\documentclass[fleqn,manuscript=article]{achemso}
\setkeys{acs}{usetitle=true} \usepackage[utf8]{inputenc} \usepackage{lineno}
\usepackage{bm} \usepackage{rotating} \usepackage[T1]{fontenc}
\usepackage{amsmath} \usepackage{amssymb} \usepackage{graphicx}
\usepackage[usenames,dvipsnames]{color}
\usepackage{comment,soul}

\usepackage[normalem]{ulem}

\usepackage{color}
\title{Surface charge pattern: impact on vibrational spectroscopy and physics of charged interfaces}

\author{Wanlin Chen}
\affiliation{Department of Physical Chemistry II, Ruhr University Bochum, D-44801 Bochum, Germany}
\author{Marie-Pierre Gaigeot}
\affiliation{Universit{\'e} Paris-Saclay, Univ Evry, CY Cergy, LAMBE UMR8587, 91025 Evry-Courcouronnes, France ; Institut Universitaire de France (IUF), 75005 Paris, France}
\author{Simone Pezzotti}
\affiliation{Laboratoire CPCV, Département de Chimie, École Normale Supérieure, PSL University, Sorbonne University, CNRS, 75005 Paris, France}
\email{simone.pezzotti@ens.psl.eu}

\date{\today}

\begin{document}

\begin{abstract}
%Charged interfaces are commonly studied using surface-specific spectroscopies, probing water’s response in the electric double layer (EDL) and correlating it with surface charge via models like Gouy-Chapman-Stern (GCS). 
Surface-specific vibrational spectroscopies revolutionized the study of charged interfaces, by sensitively probing water’s response in the electric double layer (EDL) and correlating it with surface charge via models like Gouy-Chapman-Stern. The assumed one-to-one relationship between water’s spectroscopic response and surface charge has been widely accepted without question. 
We hereby propose a theoretical experiment to evaluate this assumption. Interestingly, our findings reveal a non one-to-one relationship between surface charge and spectroscopic response, exhibiting a fascinating dependence on surface topology. 

% Surface-specific spectroscopies greatly advanced our understanding of charged interfaces in the last decades by probing water’s response within the electric double layer (EDL), which can be linked to surface charge via analytical models, e.g., Gouy-Chapman-storm (GCS). The underlying assumption of the one-to-one relationship between water’s response and surface charge has been implicitly accepted by the community. Here, we challenge this assumption by a simple theoretical experiment. Our results provide insights into the non-monogamous relationship between EDL structure and its spectroscopic response by highlighting an intriguing dependence on surface topology.

% Our results provide new insights on the non-monogamous relationship between `EDL structure and its spectroscopic response on the surface charge distribution

% Provide new insights on EDl structure and spectroscopic response by highlighting an intriguing dependence on surface topology.

\end{abstract}

\clearpage
\begin{comment}
\begin{center}
{\bf \Large Graphical TOC Entry}\\[20mm]
\includegraphics[width=0.7\textwidth]{FIGURES/GTOC.png}
\end{center}
\end{comment}

\clearpage

%\section{Introduction}

What is the protonation state of an oxide surface in contact with liquid water? What is the point of zero charge of an electrode-electrolyte interface? How do ions (and water) arrange and screen the electric field within the electric double layer (EDL)? Despite more than a century of intensive research, these fundamental questions in interface science remain challenging to address, both in theory and experiments.\cite{gonella2021water, borghetto2023oxide, gibbs2022water, Si5, koper2020pzc, levell2024emerging, gross2022ab, ringe2021implicit, wu2022understanding, cowan2019vibrational} In the last decades, surface specific spectroscopies, such as vibrational Sum Frequency Generation (vSFG) and Second Harmonic Generation (SHG), opened a possibility of gaining insights by sensitively probing the response of water within the EDL.\cite{geiger2025quantifying, xu2023optical, ong1992polarization, wen2016unveiling, gonella2021water, borghetto2023oxide, gibbs2022water, Roke_JPCC_2016, tahara2023elucidation, dalstein2023surface, dalstein2019direct, hore2024variable, gibbs2021role, Geiger_NatComm_2016, geiger2009second, roke2024does, Morita_PCCP_2018, Pezzotti_PCCP2018, Pezzotti_minerals2018, tuladhar2020ions, chen2022trail, hore2019probing, cowan2019vibrational, wang2019gate, ohno2019beyond} This response is often used to deduce the surface charge ($\sigma$), via a commonly assumed one-to-one correspondence based on the Gouy-Chapman-Stern (GCS) model and its derivations.\cite{ong1992polarization, Geiger_NatComm_2016, wen2016unveiling, Roke_JPCC_2016, Pezzotti_PCCP2018, Pezzotti_minerals2018, dalstein2019direct, ohno2019beyond}

SFG and SHG experiments probe the second order susceptibility, $\chi^{(2)}$, which is interface-specific since $\chi^{(2)} \neq 0$ only for non-centrosymmetric media, e.g. aqueous interfaces, while $\chi^{(2)} = 0$ in the centrosymmetric bulk.\cite{Shen_1994, Shen_2005, shen2008} The $\chi^{(2)}$ response from a charged interface is given by the sum of two contributions respectively arising from the topmost Binding Interfacial Layer (BIL, where the molecular organization and H-bond network substantially differ from bulk aqueous solutions due to the direct contact with the surface and ions adsorption), and from the subsequent diffuse layer (DL, composed by bulk-like water oriented by the surface electric field and by dissolved mobile ions):\cite{wen2016unveiling, Pezzotti_PCCP2018, tahara2023elucidation, gibbs2022water, gonella2021water, borghetto2023oxide, tuladhar2020ions, chen2022trail, hore2019probing, cowan2019vibrational, dalstein2019direct, backus2021probing}
 \begin{equation}
    \label{eq:chi2}
        \chi^{(2)}\left(\omega\right) = \chi_{BIL}^{(2)}\left(\omega\right) + \chi_{DL}^{(2)}
        %\chi ^{(2)}(\omega) = \chi_{BIL} ^{(2)}(\omega) + \chi_{DL} ^{(2)}(\omega)\left(\omega\right)
 \end{equation}
$\chi_{DL}^{(2)}\left(\omega\right)$ contains the response of water within the EDL. Nowadays, several approaches have been devised to extract $\chi_{DL}^{(2)}\left(\omega\right)$ from the measured $\chi^{(2)}\left(\omega\right)$,\cite{wen2016unveiling, Pezzotti_PCCP2018, tahara2023elucidation, gibbs2022water, gonella2021water, borghetto2023oxide, backus2021probing} and most recently techniques that are specifically sensitive to $\chi_{DL}^{(2)}\left(\omega\right)$, such as phase resolved SHG, have been developed.\cite{dalstein2019direct, ohno2019beyond}  As remarked in many studies, $\chi_{DL}^{(2)}\left(\omega\right)$ can be expressed as: 
\begin{equation}\label{eq:DL-chi3}
    \chi^{(2)}_{DL}(\omega) = \chi^{(3)}_{Bulk}\left(\omega\right) \int_{z_{\mathrm{a}}}^{\infty} \mathrm{d} z \cdot E_{\mathrm{DC}}(z) \mathrm{e}^{i \Delta k_z z}
\end{equation}
where $E_{\mathrm{DC}}(z)$ is the electric field profile along the z-direction perpendicular to the surface and $\Delta k_z$ is a phase factor that takes into account interferences between the emitted light at different depths from the surface.\cite{Roke_JPCC_2016, wen2016unveiling, Pezzotti_PCCP2018, Geiger_NatComm_2016, Morita_PCCP_2018, backus2021probing} The integral runs across the EDL region, with $z_a$ being the boundary between BIL and DL. $\chi^{(3)}_{Bulk}\left(\omega\right)$ is the bulk water third order susceptibility.\cite{wen2016unveiling, Pezzotti_PCCP2018} The surface charge, $\sigma$, is usually determined by using GCS to approximate the $E_{\mathrm{DC}}(z)$ dependence on z. % and solve eq.\ref{eq:DL-chi3}. 
The prerequisite of this model is that, at fixed electrolyte composition, the same $\sigma$ must always result in the same $\chi_{DL}^{(2)}\left(\omega\right)$.

Since the seminal work from Eisenthal and coworkers,\cite{ong1992polarization} this approach has been widely employed to perform SHG and SFG spectroscopic titration for oxide surfaces in contact with aqueous solutions of varying pH,\cite{ong1992polarization, tahara2023elucidation, azam2012specific, darlington2015bimodal, azam2013halide, sung2012interfacial, borghetto2023oxide} to study the way electrolyte solutions screen the surface field,\cite{ohno2019beyond, hore2014throwing, dalstein2023surface, wen2016unveiling, pezzotti2019deconvolution, tuladhar2020ions, chen2022trail, gonella2021water, borghetto2023oxide, gibbs2022water, backus2021probing, backus2022cation} to determine the charge state of the surface and the solution response in electrochemistry,\cite{geiger2025quantifying, xu2023optical, montenegro2021asymmetric, wang2019gate, cowan2019vibrational, laage2020water, liu2014situ, liu2019coherent} to probe charge separation and recombination during interfacial chemical reactions,\cite{tahara2016electron, backus2024role, backus2024ultrafast} to cite a few. Each of these studies substantially advanced our understanding of the complex chemico-physical properties of charged aqueous interfaces. However, despite being at the heart of this emerging research field, the assumption that one and only one $\chi_{DL}^{(2)}\left(\omega\right)$ corresponds to one given  $\sigma$ value (for fixed electrolyte composition) has never been assessed so far. One reason is that such an assumption is implicitly encoded into our textbook understanding of the EDL response to the electric field.

Hereafter, we test this hypothesis with a theoretical experiment, by means of classical molecular dynamics (MD) of the minimal required complexity.
%Hereafter, we present a  theoretical experiment that \red{do we want to use the word challenge here? } challenge this view. 
%\section{Results and Discussions}
%
\begin{figure}[h]
\begin{center}
\includegraphics[width=0.75\textwidth]{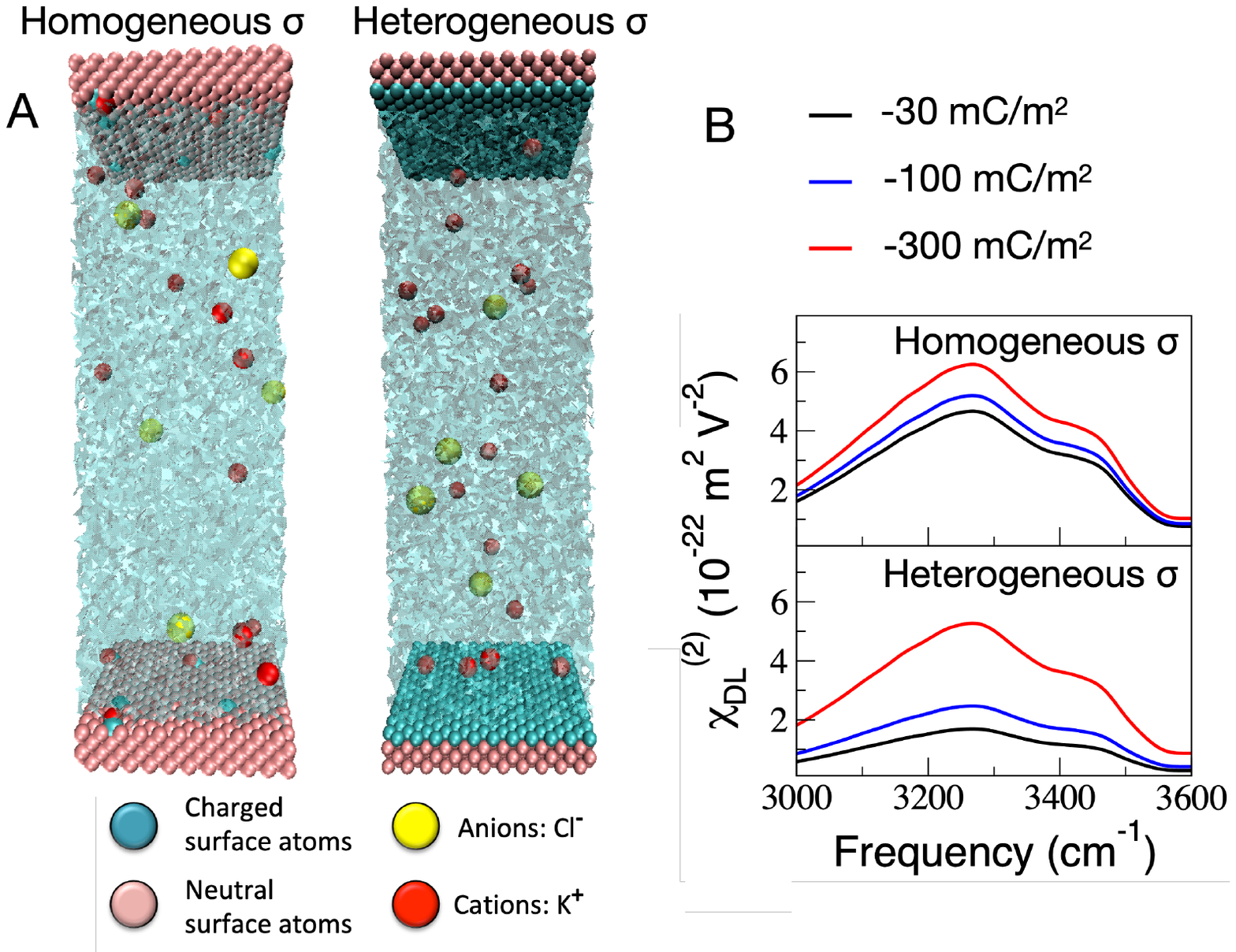}
\end{center}
\caption{Theoretical experiment challenging the commonly assumed one-to-one correspondence between $\chi_{DL}^{(2)}\left(\omega\right)$ and $\sigma$. (A) MD-snapshots illustrating the model interfaces with homogeneous and heterogeneous $\sigma$ patterns (for $\sigma$= -100 mC/m$^2$ as an example) . (B) Im$\chi_{DL}^{(2)}\left(\omega\right)$ spectra computed from  six MD simulations. The intensity is markedly different for heterogeneous $vs$ homogeneous surface charge distributions: $\sigma$ cannot be deduced from $\chi_{DL}^{(2)}\left(\omega\right)$ without knowing the surface pattern.}\label{fig:chi2-DL}
\end{figure}
%
%The experiment is designed to test our assumption with a model of the minimal required complexity, by means of classical molecular dynamics (MD). 
We used the LAMMPS~\cite{LAMMPS} code to simulate a 35 mM NaCl aqueous solution between identical charged walls  (fig. \ref{fig:chi2-DL}-A). To keep the model as simple as possible, the walls are built to represent a generic weakly interacting surface,  with the interaction parameters introduced by Huang et al.\cite{Huang2007, Huang2008} The ions concentration is kept intentionally low to ensure that we are on the range of validity of GCS, while being high enough to avoid unwanted interference effects in the $\chi^2\left(\omega\right)$.\cite{Roke_JPCC_2016} 
%This setup maximizes the success chances of the common approach based on eq.\ref{eq:DL-chi3} and GCS. 
The surface charge is implemented in two ways: (i) a homogeneous distribution among all topmost wall atoms; (ii) a heterogeneously distributed surface charge, randomly assigned to only some of the topmost atoms. A total of six model systems were considered, with $\sigma$= -30, -100, -300 mC/m$^2$, covering the typical surface charge of silica/water interfaces in the 2-14 pH range.\cite{tuladhar2020ions, tahara2023elucidation, Si5, pezzotti2019deconvolution}. Indeed, most SFG/SHG titration studies have been performed on silica/water interfaces,\cite{ong1992polarization, tahara2023elucidation, azam2012specific, darlington2015bimodal, azam2013halide, borghetto2023oxide} where SiOH terminations and therefore surface charge patterns can be more or less heterogeneous depending on the way the surface is pre-treated.\cite{rimola2013silica, cyran2019molecular, dalstein2017elusive} These studies, such as those by Gibbs {\it et al.}\cite{darlington2015bimodal, azam2013halide, gibbs2022water} and many others~\cite{review2023oxide}, have yielded some of the most intriguing yet controversial results in the field of interface spectroscopy, revealing a still-unexplained dependence of the spectroscopic response on surface preparation methods. 

We computed theoretical SFG Im$\chi_{DL}^{(2)}\left(\omega\right)$ spectra for all systems (see the method section for all details). 
\begin{figure}[h]
\begin{center}
\includegraphics[width=0.9\textwidth]{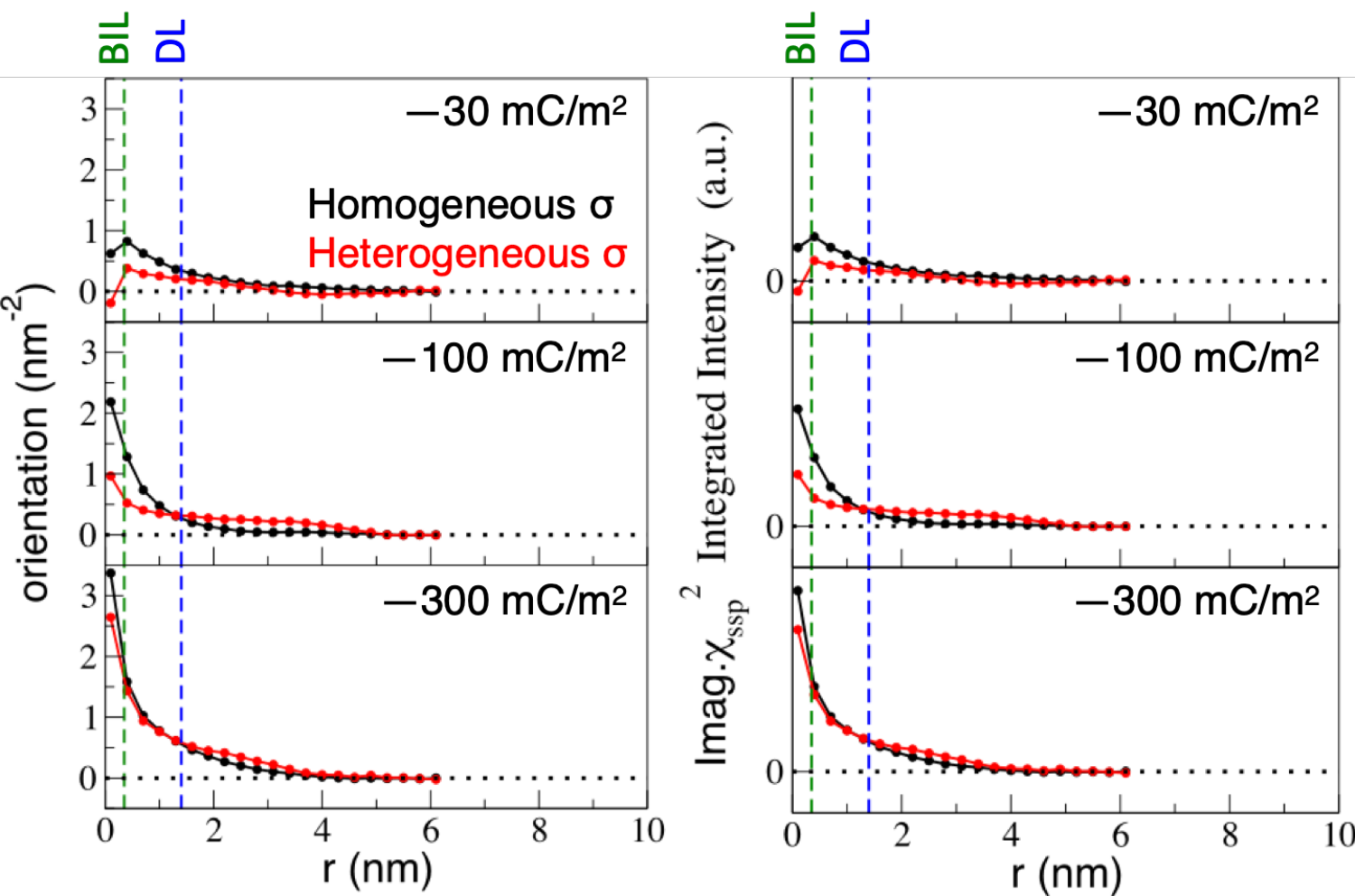}
\end{center}
\caption{Surface pattern influences the electrolyte response in the DL (diffuse layer). (A) Water orientation profile ($ \sum_{i=1}^{N{_W}} cos\theta_i$ normalized by surface area, see eq.\ref{eq:beta_eff} in the method section) along $r$ - the vertical distance from the surface. (B) Corresponding r-profiles for  Im$\chi_{DL}^{(2)}\left(\omega\right)$ intensity (integrated over 3000-3600 cm$^{-1}$), as computed from $ \sum_{i=1}^{N{_W}} cos\theta_i$ with eq.\ref{eq:beta_eff} (method section). The vertical green dashed line marks the boundary between BIL and DL regions. The blue dashed line marks a plane located 1 nm within the DL.}\label{fig:orientaion}
\end{figure}
Based on the current state of knowledge, we expect the Im$\chi_{DL}^{(2)}\left(\omega\right)$ spectrum to be identical for model interfaces with the same surface charge $\sigma$, regardless of whether the charge distribution is heterogeneous or homogeneous. However, this is not what we observe in Fig.\ref{fig:chi2-DL}-B.

While all spectra exhibit the expected two-band structure  of the Im$\chi_{DL}^{(2)}\left(\omega\right)$ (with maxima at $\sim$ 3200 and 3400 cm$^{-1}$)\cite{wen2016unveiling, Pezzotti_PCCP2018, Morita_PCCP_2018}, the intensity is strikingly higher  (by up to three times) for homogeneously charged surfaces than for their heterogeneous counterparts. The difference is most marked for the lowest $\sigma$, and slowly decreases - but persists - with increasing charging. The molecular origin of this effect lies in the sensitivity of the field-induced water alignment to the surface charge patterning, as shown in fig.\ref{fig:orientaion}-left by means of water orientation profiles. These are particularly informative since the net water orientation in the DL region is directly proportional to the intensity of $\chi_{DL} ^{(2)}$ (eq.\ref{eq:beta_eff} in the method section), as shown in fig.\ref{fig:orientaion}-right. 

All orientation profiles show the expected decay after the boundary between BIL and DL (green dashed line); the magnitude is, however, different from case to case and follows the trend observed in the topmost interfacial BIL layer. %The interfaces with a larger net orientation of DL-water are also the ones for which BIL-water better aligned. 
The orientational response of BIL water (before the green dashed line) substantially differs depending on the way the surface is charged: BIL (as well as the subsequent DL) water is systematically more efficiently aligned by the homogeneous rather than by the heterogeneous surface charge distributions (black $vs$. red in the plot). This difference should not be regarded as surprising since many recent studies, both theoretical and experimental, have demonstrated how sensitive the arrangement of hydration water is to the way the polar/charged groups are distributed over the surface.\cite{rego2022understanding, giovambattista2007hydration, lee1984structure,  xi2017hydrophobicity, rego2022learning,   wang2024topological,   Monroe_ANNREV2020, shell2024inverse, Pezzotti_JACSau2021} This is because the water response  does not depend on single water-surface interactions, but on the collective rearrangement of the water H-bond network in response to the local distribution of polar groups at the surface.\cite{rego2022understanding, xi2017hydrophobicity, rego2022learning,   wang2024topological,   Monroe_ANNREV2020, Pezzotti_JACSau2021}  In particular, studies on the wetting of patterned surfaces have shown that the surface is most wet, and accordingly, the water H-bond network is most perturbed, when the polar groups are dispersed as homogeneously as possible over the surface, while the effect is minimized when the polar groups are clustered over a few small surface areas.\cite{rego2022learning}
The same reasoning can be applied here to explain why BIL and  DL water molecules are best aligned by the homogeneous charge distributions.

In conclusion, the result of our theoretical experiment challenges the commonly assumed one-to-one correspondence between the surface charge and the electrolyte response to the electric field at the interface, as measured by $\chi_{DL}^{(2)}\left(\omega\right)$: indeed, we showed that the same $\sigma$ value can result in significantly different  $\chi_{DL} ^{(2)}$ responses depending on the specific distribution of charge across the surface. This insight may help explain the unresolved findings from the past 30 years of SFG/SHG titration studies on silica/water interfaces, where variations in surface charge patterning (i.e., the distribution of SiOH terminations) likely arose due to differences in surface preparation across studies.\cite{rimola2013silica, cyran2019molecular, dalstein2017elusive, darlington2015bimodal, azam2013halide, gibbs2022water} 
Our findings indicate that such charge patterning results in significantly different SFG/SHG intensities, as a function of pH.

More generally, our results also demonstrate that the electrolyte screening within the EDL is a 3-Dimensional (3D) problem, sensitive to the lateral charge distribution over the surface together with the specific properties of the interfacial water H-bond network. This calls for further developments of EDL theories, beyond commonly adopted 1D models (e.g., GCS and its derivatives), where surface topology and molecular interactions should be explicitly included, in line with most recent insights from MD simulation studies.\cite{willard2024electric}
We also stress the importance of independently measuring surface charge and $\chi_{DL}^{(2)}\left(\omega\right)$ to fully exploit the exciting perspectives offered by SFG and SHG spectroscopies in interface science. This is nowadays possible thanks to emerging techniques, such as the experimental scheme recently introduced by Liu and coworkers for in situ SFG spectroscopy of oxide surfaces in liquid water.\cite{Si5, liu2008sum} %Other examples are {\color{red} (for MP to write 1 sentence on  these approaches + ask the guys in germany for a reference) AFM/ATM + maybe Silvie Roke, SHS nanoparticles.}

We believe that combining SFG and SHG probes of DL water with these approaches, along with advanced theoretical molecular-level 3D EDL models,\cite{willard2024electric} will pave the way for a deeper understanding of the complex relationships between surface topology, interfacial water networks, ion distributions, and electric fields at solid-liquid interfaces.

%\section{Conclusions}
%XXX

\section{Methods}
For all simulations, the water+NaCl system was described using the force field developed by Kann and Skinner,\cite{Kann2014} with TIP4P/2005 water and rescaled (by 0.85) ions charge. Charge rescaling effectively describes ion polarisability.\cite{Kann2014}  Lorentz-Berthelot mixing rules were adopted for wall-electrolyte interactions. 
We systematically considered [NaCl] = 35 mM and added the appropriate number of sodium counter-ions to generate neutral simulation boxes, in a liquid phase of 11465 water molecule (held rigid using the SHAKE algorithm). To impose a pressure of 1 atm, we used the top wall as a piston until an equilibrium height was reached, leading to box xy-dimensions of 48.2 \AA~ and a vertical z-dimension of 267 \AA. Subsequent equilibration and production runs were of 32 ns each  (NVT ensemble, T= 298 K, Nos\'e-Hoover thermostat~\cite{Hoover_Thermostat_1985}).
 %The 32 ns simulation time is sufficient to allow the diffusion of the ions (and water) in all the simulation box.

Theoretical SFG spectra were computed following the procedure of ref.\cite{chen2023simplified}. In short, $\chi^{(2)}_{DL}(\omega)$ is expressed by the Fourier transform of the dipole ($\mu$)-polarizability ($\alpha$) correlation function,~\cite{Hynes_2002} by considering all the cross-correlation terms between the $N_{DL}$ water molecules located in the DL layer and all the $N_W$ water molecules of the simulated system:
\begin{equation}
\label{eq:DL-sum}
\chi_{DL}^{(2)}(\omega)= \frac{i\omega}
      {k_BT} \sum_{i=1}^{N{_{DL}}} \left( \sum_{j=1}^{N{_W}} \int_0^{\infty}dt e^{i\omega t}\langle \alpha_{xx}^j(t) \mu_{z}^i(0)\rangle \right) = \sum_{i=1}^{N{_{DL}}} \beta_{DL}^i(\omega)
\end{equation}
where $\beta_{DL}^i(\omega)= \frac{i\omega}{k_BT}\sum_{j=1}^{N{_W}} \int_0^{\infty}dt\exp(i\omega t)\langle \alpha_{xx}^j(t) \mu_{z}^i(0)\rangle )$ is the contribution of the i-th DL-water molecule. The SFG activity of DL-water is dominated by the dipole orientation of the water molecules,\cite{wen2016unveiling, Morita_PCCP_2018, Pezzotti_PCCP2018} 
which hence leads to  $\chi_{DL}^{(2)}(\omega)$ being calculated via the following equation:\cite{chen2023simplified}
\begin{equation}
    \label{eq:beta_eff}
    \chi^{(2)}_{DL}(\omega) = \sum_{i=1}^{N{_{DL}}} \frac{\beta_{DL}^i(\omega)}{cos\theta_i} \cdot cos\theta_i
\end{equation}
where the orientational term (with  $\theta_i$ being the angle formed between the dipole of the i-th water molecule and the normal to the surface) is directly obtained from the MD simulations, while $\frac{\beta_{DL}^i(\omega)}{cos\theta_i}$ has been parameterized in ref.\citenum{chen2023simplified}.

\begin{acknowledgement}
S. P. acknowledges funding from the European Research Council (ERC, ELECTROPHOBIC, Grant Agreement No. 101077129). W.C. acknowledges the Alexander von Humboldt Foundation (AvH) for the research fellowship under the Henriette-Hertz-Scouting Program. 
Simulations were conducted using HPC resources from GENCI-France under Grant 072484 (CINES/IDRIS/TGCC). 
We acknowledge Prof. L. Joly and Dr. Fu Li, as well as Prof. Y. Ron Shen and Prof. Weitao Liu for fruitful discussions.

\end{acknowledgement}

\end{document}